# The Physics of Life: No such thing as a dumb question

**Physics of Life research in the UK is proving to be transformative to scientific insight and translational impact, but challenges remain. Here I discuss its disruptive potential and the barriers to interdisciplinary research seen through the lens of the activities of one of its pioneers, Tom McLeish, FRS.**

Earlier this year, Tom McLeish, friend and colleague, passed away, leaving a legacy touching soft matter and polymer physics, medieval science, religion, philosophy, poetry, art and music. Here, I reflect on his impact in the Physics of Life (PoL), and the UK's Physics of Life network PoLNET.

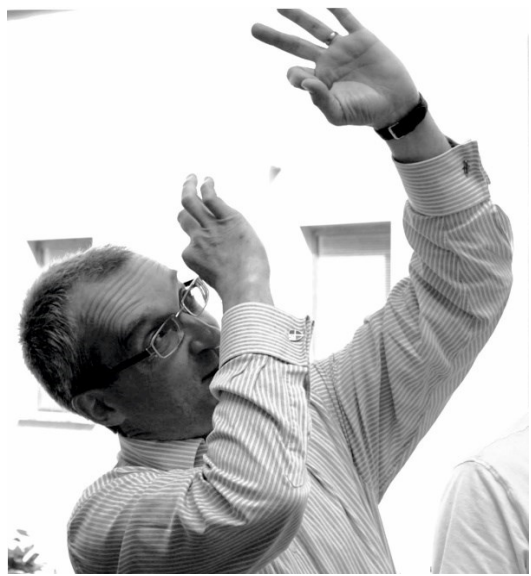

*Figure 1 Tom McLeish, FRS: 1 May 1962 - 27 February 2023. Ever a vibrant pedagogue of interdisciplinarity.*

*Biophysics* has pillars set in the foundations of 1950s physiology and structural biology[1] – *biology-led* questions requiring physics to resolve – whereas *biological physics* subsequently focused on biology as a *physics application*. More recently, a chimera emerged which sheltered both pursuits under the same cloak of the *Physics of Life*. Here, questions are not *owned* by life or physical scientists, but *shared*. This subtle distinction has transformative ramifications for how questions are contextualized, from ecosystems down to single molecules[2,3].

As part of its founder membership, I first met Tom (Fig. 1) during PoLNET's 2012 inception. PoLNET made progress quickly through targeted workshops and summer schools funded from sparse resources to encourage physical and life scientists to talk to each other. But it was during Tom's tenure as PoLNET's chair from 2017 to 2022, when the baton passed to me, that PoLNET transformed from a "convivial talking-shop" into a powerful collective for interdisciplinary research. It was down to his singular efforts that a Strategic Priorities Fund (SPF) was granted by the main UK government funder UKRI in 2018 to resource PoL research, with a budget two orders of magnitude greater than the network's itself.

**Disruptive potential**

Rather than just creating bridges between disciplines, "decorating the boundaries" as Tom would say, interdisciplinarity has a potential to enrich and transform disciplines, leading into central questions of learning[4]. A motivation for the SPF was that PoL research had enormously *disruptive* potential; to enable research by contextualizing it in a transformatively different way involving *parity* of life and physical sciences. Tom was instrumental in establishing a new structure for research consortia, insisting on "co-PI" leadership from across the phys/bio interface. This aspect of *co-creation* transformed the nature of research but also the way questions were configured.

Seeds of "joint ownership" were sown early in the life of PoLNET during open discussions, and shoots grew in fledgling local PoL groups, one such at the University of York which I established in 2015, a collective of ~50 researchers reflecting the broad national demographic of interdisciplinary expertise. When Tom moved to the University in 2018 he bought into these Friday afternoon seminar discussions and enriched them enormously. Here, we created a safe environment where enthusiastic early career researchers (ECRs) – students, postdocs, untenured fellows – mingled with tenured academics in constructive ignorance. We fostered naïve fearlessness, to not be bamboozled by expertise, but ask what, on the surface, appeared to be stupid questions. Yet it is with these apparently dumb questions posed from across the phys/bio interface that ignited the tinder for new ways of thinking about a problem. And Tom got that notion in spades, how it can result in *re-imagining the question as a dialogue*.

An interesting finding from these discussions was that this re-imagining was not solely influenced by expertise. There was also *language*, which came in two flavours. The first was *translational*; phys/bio terminology differences for the same underlying phenomena. These also occur within monodisciplines, e.g. astrophysicists describe galaxy formation as accretion differently to colloidal physicists with particle nucleation, differently to polymer physicists with phase transitions. This revelation that different parts of the same discipline are caricatures of essentially identical features can lead to rapid insights by enabling "borrowed" application from alternative contexts. Tom was a great advocate for this, exemplified by his work on liquid droplet formation in living cells by "rediscovering" Flory-Huggins theory[5] originally developed to model thermodynamics of polymer solutions. The second language factor was *epistemological*, the *structure* of knowledge; some physicists *think* of problems in different ways to biologists, and *vice versa*. So, even if the objects of knowledge are the same, the nature of phenomena involving them might be differently perceived because the connecting *wiring* is discipline-specific. Recognizing that different arrangements exist poses opportunities to develop new *idea architectures* that are neither homed in either physics or biology, and indeed can be seen reflected in much of the SPF-funded PoL research across the UK[6,7].

**Interdisciplinarity challenges**

Recent reports highlight the power and tribulations of UK interdisciplinarity, concerning research councils [8], the research excellence framework (REF) [9], cultural challenges within UK academia[10], and housing PoL in physics departments[11]. A common observation was that interdisciplinarity taps into UK funder remits well but is hampered by organisational/administrative structures. Some of these barriers are beginning to erode through merging councils into a single UKRI umbrella, but many remain. Challenges of interdisciplinary funding and career sustainability were reflected in a straw poll I took while chairing a recent PoL roadmap discussion[12]; this event was a legacy of Tom's impact in PoL – a unique meeting in which major UK stakeholders of PoL/biological physics/biophysics research came together. ~70% of participants identified as ECRs, ~1/3 had core background in life and physical sciences each and the remaining ~1/3 claimed equal expertise in both; when asked what they would most like to discuss, "funding" and "career" had prominence (Fig. 2).

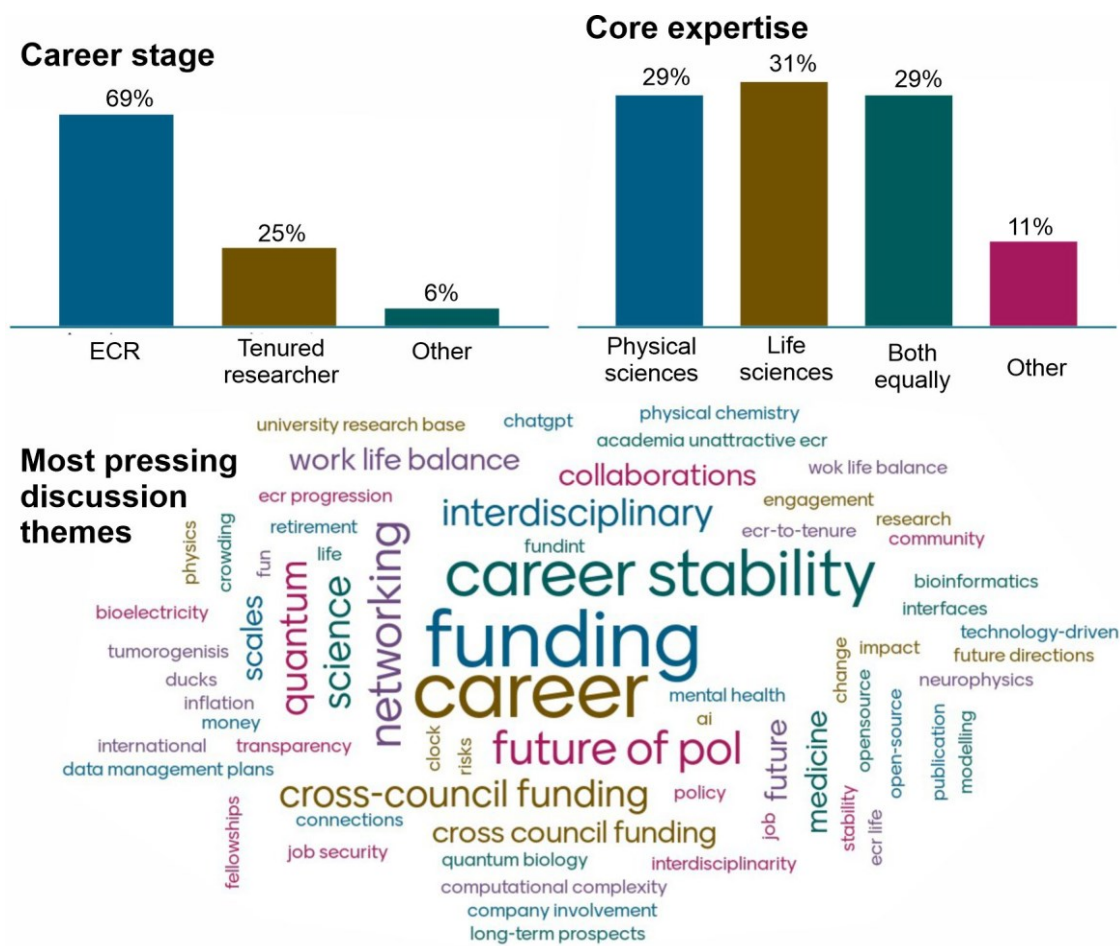

*Figure 2 Mentimeter.com poll taken during Physics of Life Roadmap discussion panel at PoL 2023 conference, Harrogate, UK, 29 March 2023 , involving ~200 respondents to questions concerning career stage, area of self-indentified research, and what they would most like to discuss.*

Finding that the bulk of participants were ECRs was no surprise, but the balance between life and physical science, and the presence of a significant population identifying equally with both, was; at PoLNET's inception, a barrier was getting life and physical scientists in the same room at the same time. But, due to PoLNET, this colocation challenge is now significantly diminished.

There were anecdotal suggestions that inappropriate reviewer expertise of PoL grants was still a challenge, however, this problem appeared less prevalent than during initial discussions a decade ago. What *was* flagged was the challenge of judging grants by monodisciplinary funding panels, to ensure fairness when PoL and monodisciplinary bids compete. However, an important success of the PoL SPF process was the establishment of a cross-disciplinary panel; to roll such out as *business as normal* could have major benefits.

A larger issue regarding how PoL grants are assessed was a perception that PoL reviewers are more conservative in rating their peers' research compared to monodisciplines. Do some biologists more harshly evaluate proposals of physicists, and *vice versa*? Do some researchers sternly reviewed previously reflect indignation when evaluating others? Is the PoL community "immature" compared to disciplines like nuclear and quantum physics whose members score more favourably when reviewing peers? These speculations need to be further debriefed through open community discussion. However, although there are aspects of getting one's house in order that are worth reflecting on for PoL, a more pressing aspect is how best to resource research, which requires discussion with those holding the purse strings.

**Physics of Life roadmap**

UK PoL research is at a turning point with more funding needing to be secured, exemplified by the submissions of grants to the most recent PoL SPF call: 171 outline bids submitted but resources sufficient to fund only nine. In having been privy to successful and unsuccessful bids I sense in cases only marginal distinctions discriminating triumph from failure; the quality was there in abundance. £33M sounds plenty, however the UKRI contribution is equivalent to 1% of its total budget for the financial years 2022-25. As a "cross-cutting" activity spanning multiple branches of the UKRI tree which transformatively resolves societal challenges through innovative research, is 1% enough?

An important factor for PoL's future lies with nurturing support from charities which sponsor interdisciplinarity, especially at the physics/biomedicine borders. UKRI supports research echoed, at the time of writing, in five strategic priorities[13] – building a green future, a secure/resilient world, creating opportunities/improving outcomes, securing better health/ageing/wellbeing, tackling infections. If the PoL community works with UKRI to align with these this will surely favour increased resourcing; there are indicators of existing alignment in recent funded PoL themes: tissue morphogenesis, brain function, plant resilience, understanding super-bugs, innovations to capture carbon inspired by simple algae... Increasing scope for themed challenges in PoL research from multiple bodies across physical sciences, biology, biomedicine and the environment, will serve to diversify funding, increasing resilience when individual funders tighten belts. There are also funding lessons to learn from international partners, valuable recommendations worth musing over in the US National Science Foundation recently published an extensive report from its Physics of Living Systems network into future challenges[14], and also looking towards alternative localised hub models for theming interdisciplinary research as seen in many collaborations between geographically close universities of the Netherlands, and in the >80 Max Planck institutes of Germany.

Increased engagement involving non-academic partners also presents opportunities for PoL. As Tom pointed out regarding soft-matter physics "[it] does exactly what it says on the tin… and very often it is in the tin…" – the implication being that it has real-world implications manifest in industries ranging from food production to shampoo design, that he recognized were desperate for scientists skilled in physical *and* life sciences. But it is crucial to recognize the effort involved in negotiating *challenge-led* projects that balance industry needs against basic science development. UKRI facilitates Prosperity Partnerships and iCASE PhD studentships to catalyze industrial engagement, but a challenge remains in the disparity between the few large multinationals that can invest in these initiatives versus the larger number of small and medium-sized enterprises (SMEs) who have interests in engagement but struggle to commit major resourcing. It is vital with industrial/academic engagement to balance translation with discovery, but the importance of *curiosity*, of *asking dumb questions*, should never be forgotten; the science successes in UK's COVID19 vaccination response illustrate this, since much of our readiness emerged following many years of curiosity-driven research.

One clear issue which a PoL roadmap must address is how to grow its pool of ECRs, and to ensure that we cultivate the community's diversity, an area very close to Tom's heart from his chairing of the Royal Society's Education Committee. The traditional model of sponsoring young researchers via tenure track programmes and independent fellowships works for the gifted few; it is imperative that funders of excellence in interdisciplinarity continue to do so. But a challenge remains: these traditional routes can unintentionally exclude on the basis of gender, social status, age and geography, and simply do not cater well for many postdocs who yet still have enormous expertise. ECR mentorship is vital here, and PoLNET is playing an important role in coordinating this. One further suggestion is to create more *research officer* (RO) positions; senior postdoctoral posts supporting multiple teams whilst still enabling a path to independence. In response, PoLNET is working with UKRI to develop ECR "pump-priming" to better nurture the next generation of PoL researchers. Increased outreach into schools will also be invaluable – the dye is currently cast early in the learning journeys of our schoolchildren between the life and physical sciences, which has frustrating implications for diversity in the physical sciences in particular at university level. A really essential action for growing the PoL community is to continue to support the underpinning activities of PoLNET itself – it will likely have a pivotal role in securing that common vision and purpose that will be enormously beneficial for the UK.

**Reflections**

I will miss Tom enormously, his intelligence, humor and kindness. He understood that researchers and decision-makers are human, often resorting to traditional temptations of monodisciplinary research. But what he also understood was never to give up on anybody, to listen to their needs and fears, to help reveal the power of research with non-traditional colleagues; having faith to take your hands off the reins for a while and just allow the horse to do the steering.

His passing leaves a void, but one being filled by the emergence of inspirational research in PoL, driven by the next generation of ECRs. I feel blessed to have known Tom, and to have played a small part with him in search of understanding through asking what appear to be dumb questions. PoL all starts with that. For, as Tom knew very well, there actually really is no such thing as a dumb question.

*Mark Leake is Coordinator of the Physics of Life Group at the University of York, and serves as Chair of the Biological Physics Group of the Institute of Physics, and for the UK Physics of Life network PoLNET.*

**Mark C. Leake**[1,2]

[1] School of Physics, Engineering & Technology, and [2] Department of Biology, University of York, York YO10 5DD, UK.

email: mark.leake@york.ac.uk

**Acknowledgements**

The author is Chair of the UK Physics of Life Network (PoLNET) and acknowledges the enormous efforts from its steering group members, past and present, since its inception in 2012, its flourishing membership of over 1,000 researchers from all career levels spanning experiment and theory and multiscale expertise from the physical and life sciences, and the UKRI funding of its activities (EP/T022000/1). Thanks to Prof Giles Gasper (University of Durham) for permission to use the photograph of Tom McLeish (Fig. 1).